\documentclass[%
 reprint,
 amsmath,amssymb,
 aps,
 showkeys,
]{revtex4-2}

\usepackage{graphicx}
\usepackage{dcolumn}
\usepackage{bm}
\usepackage{booktabs}
\usepackage{xurl}
\usepackage{hyperref}

\begin{document}

\renewcommand{\thesection}{\arabic{section}}
\renewcommand{\thesubsection}{\thesection.\arabic{subsection}}


\title{Planetary Surface Temperatures from First Principles: Geometric Insights into Energy Balance and Implications for Habitable Exoplanets}

\author{Sabin Roman}
\email{Sabin.Roman@ijs.si}
\affiliation{Department of Knowledge Technologies, Jo\v{z}ef Stefan Institute, 1000 Ljubljana, Slovenia}


\begin{abstract}
A previously overlooked relation governing planetary surface temperatures in terms of solar irradiance and top-of-atmosphere Bond albedo is identified. It reproduces the observed climates of Venus, Earth, and Titan, predicts condensation-level temperatures in the gas giants Jupiter, Saturn, Uranus, and Neptune, and extends naturally to rocky planets and large moons with substantial atmospheres. The relation encodes global energy conservation and highlights Bond albedo as a key bulk radiative constraint. Its central result is an empirical proportionality between Bond albedo and the fraction of outgoing longwave radiation returned downward by the atmosphere, termed the inner albedo. A geometric argument based on local beam-aligned parabolic wavefronts provides a rationale for a coefficient linked to the parabolic constant. Compared with classical one- or multi-layer models, the formulation achieves strong agreement using directly measurable quantities and no planet-by-planet tuning. Applied to exoplanets, it yields first-order estimates of equilibrium surface conditions across the habitable zone, suggesting that a substantial part of planetary temperature structure may be constrained by a simple relation among bulk radiative observables.
\end{abstract}

\maketitle

\section*{Research Highlights}
\begin{itemize}
\item Relation governing planetary surface temperatures in terms of solar irradiance and top-of-atmosphere Bond albedo.
\item The relation encodes global energy conservation and highlights Bond albedo as a key bulk radiative constraint.
\item Applied to exoplanets, first-order estimates of equilibrium surface conditions across the habitable zone.
\end{itemize}


\section{Introduction}
\small
Beyond their empirical successes, climate models raise a foundational question: to what extent can planetary climates be explained by physical invariants, rather than by a proliferation of process-specific parameters? This paper addresses that question. Consider a planet of radius $R$ receiving incident stellar radiation characterized by the solar constant $S$, i.e., the flux of energy per unit area at the planet’s orbital distance. The total intercepted power is then $\pi R^{2}S$, corresponding to the cross-sectional area of the planet times the incident flux. Averaged over the full planetary surface area, $4\pi R^{2}$, the mean incoming irradiance is $S/4$. We denote this quantity as the average solar irradiance $I$. For Earth, where $S = 1361 \,\text{W m}^{-2}$, this yields $I = 340.2 \,\text{W m}^{-2}$ (see Table \ref{table:1}).

A portion of the incident radiation is reflected back into space, quantified by the top-of-atmosphere Bond albedo $\alpha$, which throughout this paper denotes the planetary Bond albedo relevant for the global energy balance. The effective absorbed flux is therefore $I(1-\alpha)$. For Earth, with $\alpha \approx 0.30$ \cite{NASAfacts,stephens2015albedo}, the average absorbed flux is about $236 \,\text{W m}^{-2}$. The planetary surface temperature may then be estimated using the Stefan–Boltzmann law, which states that a body at temperature $T$ radiates an energy flux $\sigma T^{4}$, where $\sigma$ is the Stefan–Boltzmann constant. Setting $I(1-\alpha) = \sigma T^{4}$ gives the familiar blackbody equilibrium temperature. For Earth this yields $T_{0} \approx 254 \,\text{K}$ ($-19\,^{\circ}\text{C}$) \citep{NASAfacts}, which is significantly below the observed mean surface temperature of $T_{p} \approx 288 \,\text{K}$ ($15\,^{\circ}\text{C}$), an offset of roughly $34 \,\text{K}$.

This discrepancy reflects the limitations of the blackbody approximation, which neglects the role of the atmosphere in trapping and re-emitting longwave radiation. A useful way to quantify this effect is through the ratio of absorbed solar flux to outgoing surface radiation, $I(1-\alpha)/\sigma T_{p}^{4} \leq 1$.

\begin{figure}[t]
\centering
\includegraphics[width=0.5\textwidth]{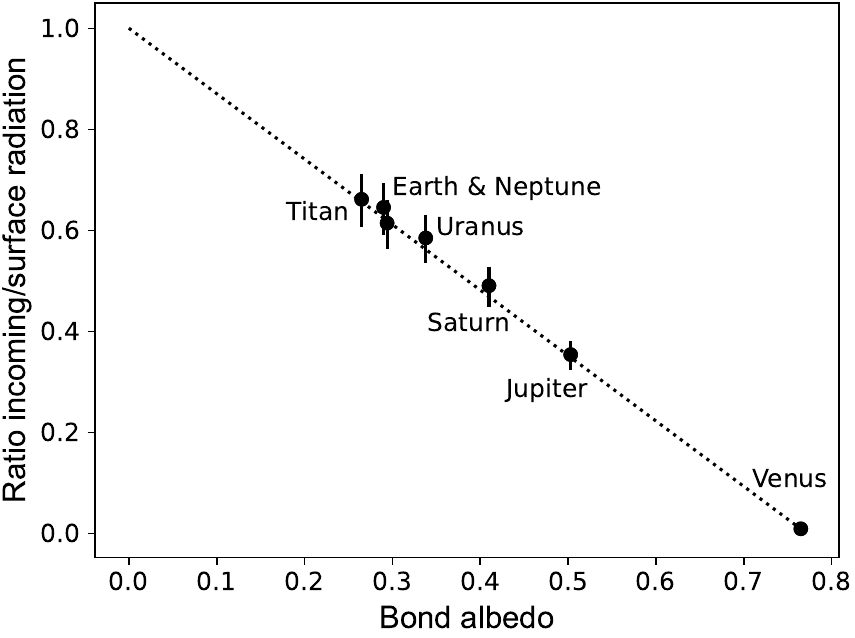}
\caption{Ratio of absorbed solar irradiance to surface-emitted radiation, plotted against the Bond albedo for Titan, Earth, and Venus. Gas giants are included by assigning the ``surface'' to the atmospheric level where condensates first form. Values are taken from Table \ref{table:1}, with 2\% error bars reflecting uncertainties in temperature estimates. The linear relation demonstrates a systematic dependence of radiative balance on albedo.}
\label{fig:1}
\end{figure}

When this ratio is computed for Titan, Earth, and Venus, and plotted against their respective Bond albedos, a striking linear relationship emerges (Figure~\ref{fig:1}). A least-squares fit gives
\begin{equation}
\frac{I(1-\alpha)}{\sigma T_{p}^{4}} = 1 - r\alpha,
\label{eq:1}
\end{equation}
with slope $r \simeq 1.295(15)$. Defining the emitted surface flux as $I_{p} = \sigma T_{p}^{4}$ and introducing $\beta = r\alpha$, Equation~\eqref{eq:1} can be recast as
\begin{equation}
I(1-\alpha) = I_{p}(1-\beta).
\label{eq:cons}
\end{equation}

\begin{table*}[t]
\centering
\caption{\label{table:1}
Predicted planetary surface (or reference-level) temperatures using Equation~\eqref{eq:1} with
$r=\log(1+\sqrt{2})+\sqrt{2}-1$. Bond albedos from: Venus \citep{NASAfacts,haus2016radiative}, Earth \citep{stephens2015albedo},
Titan \citep{creecy2021titan}, Jupiter \citep{li2018less}, Saturn \citep{hanel1983albedo,wang2024saturn}, Uranus \citep{irwin2025uranus},
Neptune \citep{irwin2022hazy,pearl1991neptune}. Observed temperatures from: Venus \citep{NASAfacts,seiff1985models}, Earth \citep{Hartmann2016Climate},
Titan \citep{mckay1991greenhouse,jennings2016surface}, Jupiter \citep{seiff1998thermal}, Saturn \citep{lindal1985atmosphere},
Uranus \citep{lindal1987atmosphere,lunine1993atmospheres}, Neptune \citep{lindal1990atmosphere,fletcher2010neptune}.
Solar irradiances are from NASA Planetary Fact Sheets \citep{NASAfacts}. Prediction uncertainties are 1$\sigma$ propagated from albedo uncertainty,
except for Venus where no uncertainty is shown due to sensitivity near the model bound.}
\setlength{\tabcolsep}{12pt}
\renewcommand{\arraystretch}{1.2}
\begin{tabular*}{\textwidth}{@{\extracolsep{\fill}} l c c c c}
\toprule
Planet & Average Solar Irradiance\,(W\,m$^{-2}$) & Bond Albedo $\alpha$ & Observed $T_{\mathrm{planet}}$ (K) & Predicted $T$ (K) \\
\midrule
Venus   & 650.3 & 0.765                 & 737          & 742.3 \\
Earth   & 340.2 & $0.294 \pm 0.005$     & 288.2        & $287.6 \pm 0.2$ \\
Titan   &   3.7 & $0.265 \pm 0.003$     & 90.6--94     & $92.5 \pm 0.04$ \\
Jupiter &  12.6 & $0.503 \pm 0.012$     & 132.8        & $133.3 \pm 0.7$ \\
Saturn  &   3.7 & $0.410 \pm 0.020$     & 93.5--94.8   & $95.2 \pm 0.5$ \\
Uranus  &  0.923& $0.338 \pm 0.011$     & 64.4--66.6   & $66.2 \pm 0.1$ \\
Neptune &  0.377& $0.290 \pm 0.067$     & 51--53       & $52.4 \pm 0.6$ \\
\bottomrule
\end{tabular*}
\end{table*}

Equation \eqref{eq:cons} admits a natural interpretation in terms of energy balance: the effective incoming flux is $I(1-\alpha)$, while the net outgoing flux is reduced to $I_{p}(1-\beta)$. The parameter $\beta$ may be interpreted as an effective ``inner albedo,'' denoting the fraction of upwelling thermal radiation that is returned downward by the atmosphere. Because $\alpha$ and $\beta$ are linked in the present framework, they may be regarded as coupled radiative effects of the same large-scale atmospheric structures, although they arise from different shortwave and longwave transfer processes. In the limiting case $\beta = 0$, atmospheric effects vanish and the relation reduces to the blackbody estimate, which systematically underpredicts observed temperatures. While sunlight can be treated as nearly parallel rays because of the large distance between the planet and the star, surface-emitted radiation originates locally and therefore interacts with the atmosphere under a different geometric configuration. If $\beta$ is assumed to be linked to $\alpha$, a naive first guess would be $\beta = \alpha$. However, that identification would implicitly treat the upwelling thermal flux as geometrically analogous to the incident solar beam, which it is not. Instead, the proximity of the emitting surface introduces a geometric enhancement of the inner albedo, a point developed in the next section.

Solving Equation~\eqref{eq:1} for $T_{p}$, we obtain
\begin{equation}
T_{p} = \left[\cfrac{I(1-\alpha)}{\sigma (1-r\alpha)}\right]^{1/4} = \cfrac{T_{0}}{(1-r\alpha)^{1/4}},
\label{eq:pred}
\end{equation}
where $T_{0}$ is the black-body estimate. For any given planet, Equation~\eqref{eq:pred} can trivially be made to reproduce the observed temperature by choosing $r$ as a fitted parameter. In that sense, the relation is not by itself restrictive if $r$ is allowed to vary freely between planets.

\begin{figure}[t]
\centering
\includegraphics[width=0.5\textwidth]{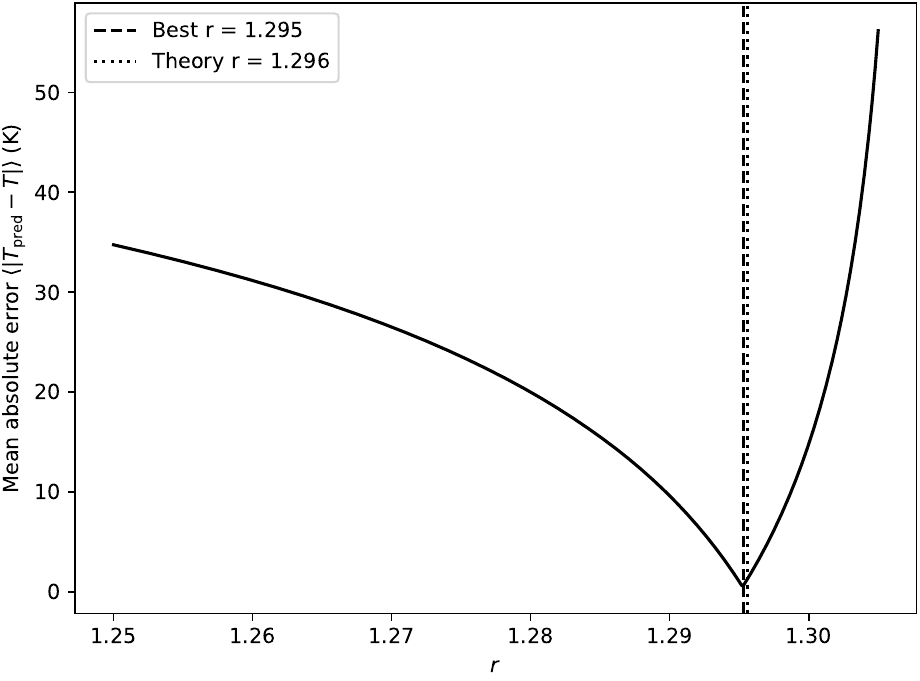}
\caption{Mean absolute error between predicted and observed planetary temperatures as a function of $r$. The minimum occurs near $r \approx 1.295$, consistent with the theoretical value.}
\label{fig:min_r}
\end{figure}

The nontrivial result is that a single, universal value of $r$ provides accurate predictions across diverse planetary environments. The linear relationship in Figure~\ref{fig:1} and the error analysis in Figure~\ref{fig:min_r} show that the mean absolute error exhibits a well-defined minimum at $r \simeq 1.295$, in close agreement with the theoretical value $r = \log(1+\sqrt{2})+\sqrt{2}-1 \simeq 1.296$. With this single value, Equation~\eqref{eq:pred} reproduces the observed temperatures of Venus, Earth, Titan, and the giant planets (at their reference levels) within observational uncertainty. This suggests that $r$ is not an adjustable parameter but reflects a common structural feature of planetary energy balance.

The expression also implies a limiting albedo $\alpha^\star = 1/r$, beyond which the denominator vanishes and the model ceases to be well-defined. Venus lies close to this bound, which accounts for both its high surface temperature and the strong sensitivity of the prediction to small variations in $\alpha$. Since the denominator in Equation~\eqref{eq:pred} is generally less than one, the predicted temperature exceeds the black-body value, as expected for atmospheres that return a fraction of outgoing longwave radiation.

The appearance of a universal value of $r$ points toward a geometric origin, developed in Section 2, that is independent of the detailed microphysics of individual atmospheres. This stands in contrast to traditional effective models, where atmospheric effects are introduced through fitted parameters.

A classical example is the one-layer atmospheric model, which predicts
\begin{equation}
T_{1} = \frac{T_0}{(1-\epsilon/2)^{1/4}},
\label{eq:layer}
\end{equation}
where $T_0$ is the black-body temperature and $\epsilon$ the effective longwave emissivity \citep{hansen1981climate}. Models of this type represent the simplest effective descriptions of atmospheric radiative dynamics and are often used as starting points for more elaborate radiative--convective frameworks. In that sense, they provide the most direct point of comparison to the present approach.

In practice, however, the one-layer model relies on a fitted emissivity to reproduce the observed surface temperature. For Earth, the value $\epsilon/2 \simeq 0.39$ coincides numerically with $\beta = r\alpha$, but $\epsilon$ must be tuned to match observations on a case-by-case basis, whereas $\beta$ is directly determined by $\alpha$. Moreover, because $\epsilon < 1$ by construction, the one-layer model cannot reproduce high-temperature cases such as Venus. In such regimes, additional layers must be introduced, effectively adding further adjustable parameters (e.g., the number of layers), so that the model becomes a multi-parameter fitting framework.

Both Equation~\eqref{eq:1} and Equation~\eqref{eq:layer} express global energy conservation, but they differ in how the atmospheric contribution is represented. The one-layer formulation requires, in general, the irradiance $I$, the Bond albedo $\alpha$, and one or more fitted parameters. By contrast, Equation~\eqref{eq:pred} uses only directly measurable quantities, $I$ and $\alpha$, to determine the surface or reference-level temperature.

From an effective or coarse-grained perspective, this makes Equation~\eqref{eq:1} a more constrained and predictive starting point. Rather than introducing tunable parameters to represent atmospheric complexity, it captures the dominant radiative coupling through a single observable quantity, the Bond albedo, and its geometric linkage to longwave return.

In Section 2 we propose a geometric rationale for the universality of $r$, while Section 3 explores applications to the gas giants and exoplanets. For the gas giants, the same relation appears to capture the temperature at which condensates first form in their atmospheres, marking an outer radiative--convective transition region shaped by solar forcing, even though the deeper atmosphere is also influenced by intrinsic heat flux. To illustrate this broader applicability, Jupiter, Saturn, Uranus, and Neptune are also included in Figure~\ref{fig:1}, though the linear fit itself is based only on the rocky bodies listed in Table~\ref{table:1}.

\section{Model of Inner Albedo}
\label{sec:geometry}

Given the large differences in atmospheric composition, surface features, and other characteristics between Venus, Earth and Titan, the fact that Equation~\eqref{eq:1} holds empirically so well suggests that $r$ is geometric in nature. In this section we propose a model to account for Equation~\eqref{eq:1}, which can justify the universal nature of $r$.

\begin{figure}[t]
\centering
\includegraphics[width=0.5\textwidth]{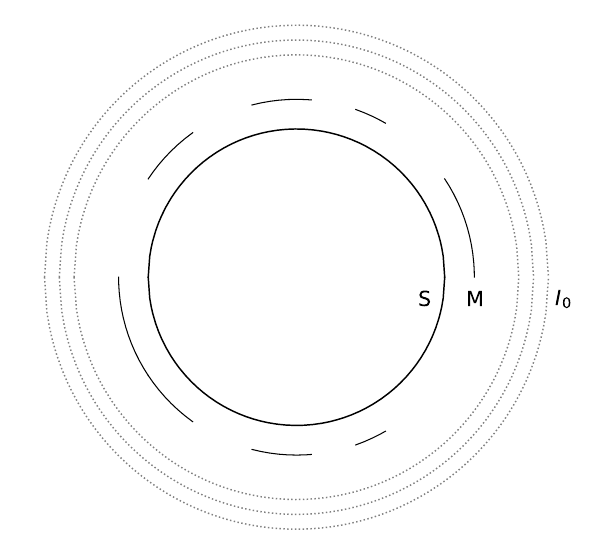}
\caption{Incoming light of irradiance $I_{0}$, passing through the transparent segments $M$, reflecting off the rough surface $S$ and reflecting again off the inner surface of $M$.}
\label{fig:2}
\end{figure}

Consider a reflective sphere $S$ with a rough surface such that light scatters in the outward tangent half-space at every point. We make two main simplifying assumptions, which will be discussed at the end of the section. The first assumption is that we restrict our analysis to a two dimensional cross-section made by a plane through the centre of $S$, see Figure~\ref{fig:2}. The cross-sectional circle is surrounded by a set $M$ of circular segments that are transparent to incoming light but reflective on their inner surface. Furthermore, the mirrors $M$ are at a close distance from the surface relative to the radius of the sphere $S$. The second key assumption is that when the light reflects off $S$ it does so with a wavefront well-approximated by the parabola $P$ with focus $F$ from Figure~\ref{fig:3} (see Appendix A for a detailed discussion). We are interested in how the light reflects from the inner surface of $M$, and we examine segment $AB$ of $M$ in Figure~\ref{fig:3}.

We distinguish three intensities in the local construction: $I_{0}$ is the externally incident intensity passing through the transparent part of $M$, $I_{2}$ is the intensity re-emitted upward from the surface $S$, and $I_{1}$ is the intensity returned downward by the reflective segment $M$ after the first reflection. 

\begin{figure*}[t]
\centering
\includegraphics[width=0.9\textwidth]{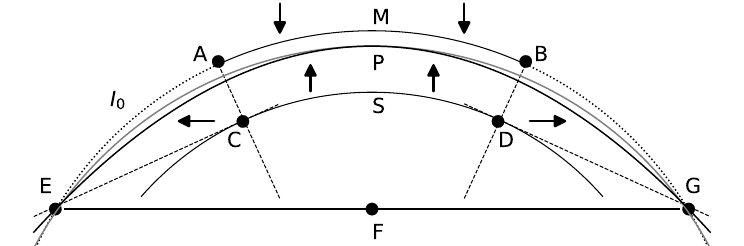}
\caption{The geometry around a segment $AB$ of $M$. Lines $AC$ and $BD$ go through the origin, which is the centre of $S$. Tangents at $C$ and $D$ intersect the circle encompassing $AB$ at points $E$ and $G$. The midpoint of $EG$ is the focus $F$ of the parabola $P$. Light comes in with irradiance $I_{0}$ (downward arrows), reflects off $M$ with exitance $I_{1}$, and again off $S$ with exitance $I_{2}$ (upward and side arrows).}
\label{fig:3}
\end{figure*}

Let $l_{1}$ be the length of the circle segment $AB$ and $L_{1}$ be the length of the parabolic segment $EG$.  First, the external radiation with intensity $I_{0}$ reaches the surface through the transparent part of $M$. On the first pass, this incident input provides the source for the surface re-emission, so $I_{2}$ is identified with $I_{0}$ at that stage. Second, the reflected component $I_{1}$ generated by $M$ adds to the incoming radiation and therefore contributes to subsequent surface re-emission. Third, because the same geometric relation applies at each later step, the ratio of returned intensity to upwelling surface intensity remains the same throughout the iteration, or equivalently converges to the same value in equilibrium. Thus the ratio obtained from the first pass also determines the effective equilibrium return fraction.

With this interpretation, conservation of energy in the two-dimensional setting gives the relation between the incoming light $I_{0}+I_{1}$ (the externally incident component plus the returned component) and the radiation re-emitted from the surface with intensity $I_{2}$:
\begin{equation}
l_{1}(I_{1}+I_{0}) = L_{1} I_{2}.
\end{equation}
On the first pass we identify the emitted surface intensity with the incident one, so that $I_{2}=I_{0}$. Substituting this into the conservation law gives
\begin{equation}
I_{1} = \left(\frac{L_{1}}{l_{1}}-1\right)I_{0}.
\end{equation}
Hence the ratio $I_{1}/I_{0}$ is fixed entirely by geometry.

From the geometry in Figure~\ref{fig:2} we can deduce the approximate equality $FG \simeq AB \simeq l_{1}$, valid in the limit $h \ll AB \ll R$ (see Appendix A). The ratio of the parabolic segment $EG$ of length $L_{1}$ to the linear segment $FG \simeq l_{1}$ is given by the universal parabolic constant $p = \log{(1+\sqrt{2})}+\sqrt{2} \simeq 2.2956$. Therefore,
\begin{equation}
\frac{I_{1}}{I_{0}} = p-1.
\end{equation}

We now extend the local relation to the full effective circumference. Let $\sum_{i} l_{i}$ be the summed lengths of all the segments of $M$, and let $C$ be the circumference of the circle. By definition, these segments occupy a fraction $\alpha$ of the circumference, so that
\begin{equation}
\frac{\sum_{i} l_{i}}{C} = \alpha.
\end{equation}
The effective inner albedo $\beta$ is defined as the fraction of the upwelling surface radiation that is returned downward by the atmospheric segments. In the present construction, this is obtained by comparing the total returned intensity from the segments of $M$ with the total emitted intensity from the surface along the effective circumference. Since the same return ratio applies on the first pass and is preserved under subsequent reflections, the equilibrium value of $\beta$ is
\begin{equation}
\beta
=
\frac{I_{1}\sum_{i} l_{i}}{I_{0} C}
=
\left(\frac{I_{1}}{I_{0}}\right)\left(\frac{\sum_{i} l_{i}}{C}\right)
=
(p-1)\alpha.
\end{equation}
Thus, we can consider the inner albedo of $M$ to be $\beta = r\alpha$ where $r = p-1 \simeq 1.2956$. This theoretical value gives a close match of temperature predictions with data, see Table~\ref{table:1} and Figure~\ref{fig:min_r}.

The above analysis made two key simplifying assumptions: (1) a restriction to a two-dimensional setting, and (2) a parabolic approximation for the wavefront of radiation emitted from the surface $S$. Regarding assumption (2), note that the parabola $P$ in Figure~\ref{fig:3} approximates the corresponding circular wavefront, represented schematically by the grey curve. Huygens' principle is therefore not violated. Rather, as discussed in Appendix~A, the parabola is the natural local beam-aligned representation of the outgoing front.

Regarding assumption (1), the set $M$ denotes the effective atmospheric segments intercepted by the chosen two-dimensional cross-section. These include clouds, but are not restricted to them; more generally, $M$ represents the radiatively relevant atmospheric structures that contribute to the return flux. Because such structures are generally irregular in shape and distribution, they do not naturally support a rotationally symmetric idealization. This is why the relevant local geometry is taken to be a beam-aligned cross-section rather than a radially symmetric three-dimensional construction. The segments of $M$ are taken to be transparent to the incoming radiation and reflective on their inner side for the returning component. Appendix~A shows that this two-dimensional construction extends consistently to three dimensions through a decomposition into thin beam-aligned slabs. In that setting, the relevant local fronts are narrow parabolic cylinders obtained by extending the beam-aligned parabolic cross-section over a small azimuthal width. The two-dimensional argument used here is therefore the thin-slab reduction of the corresponding three-dimensional geometry.

The argument we have presented is geometric in nature and applies to rocky celestial bodies such as Venus, Earth and Titan. In the next section we discuss extensions to gas giants and exoplanets.

\section{Discussion and Applications}

The results in  Table~\ref{table:1} and the preceding derivation show that planetary surface temperatures are governed by a simple invariant closure. If this relation is correct, then any detailed radiative–convective or circulation model must either incorporate it explicitly or reduce to it in the appropriate limit. This absence in climate modeling highlights the need for a fundamental re-examination of its theoretical foundations.

Equation~\eqref{eq:1} may be interpreted in terms of an effective thermal mirror mechanism for atmospheres containing optically thick cloud or haze layers.  Any optically thick atmospheric layer in local thermodynamic equilibrium absorbs upwelling thermal radiation and re-emits it nearly instantaneously, as its heat capacity is negligible. We denote by $\beta$ the fraction of upwelling longwave flux that is absorbed and re-emitted downward by the atmosphere. The corresponding shortwave quantity is the top-of-atmosphere Bond albedo $\alpha$, namely the fraction of incident stellar radiation reflected back to space, which generally includes contributions from clouds, hazes, atmospheric scattering, aerosols, and surface reflection as mediated by the atmosphere. In atmospheres where cloud or haze layers dominate the radiative architecture, the same large-scale structures that enhance TOA shortwave reflection can also enhance longwave radiative return. Motivated by the geometric argument of Section~\ref{sec:geometry}, we therefore write
\[
\beta \;=\; r\,\alpha, \qquad r=p-1=\log(1+\sqrt{2})+\sqrt{2}-1 \approx 1.296.
\]
This proportionality is intended as a coarse-grained hypothesis: it does not assert identical shortwave and longwave microphysics, but treats their net effects as coupled through the same large-scale atmospheric structures. In that sense, cloud and haze layers are not the sole determinants of planetary albedo, but in cloud-rich regimes they can provide the dominant macroscopic link between shortwave reflection and longwave return, consistent with longstanding discussions of the central role of clouds in climate sensitivity and radiative complexity \citep{hansen1981climate,rind1999complexity,wallace2006atmospheric}.

The physical picture is then as follows. Absorbed stellar radiation warms the lower boundary of the radiating system, typically the surface for terrestrial planets or deeper atmospheric layers for the giant planets. That energy is returned upward as thermal radiation, a fraction of which is absorbed and re-emitted downward by the atmosphere; in the present closure this return flux is quantified by $\beta=r\alpha$. The system warms until the outgoing flux to space balances the absorbed stellar input, consistent with Equation~\eqref{eq:1}. The proportionality between $\beta$ and the TOA Bond albedo $\alpha$ therefore expresses an effective linkage between shortwave reflection and longwave radiative return by the same large-scale atmospheric structures. This interpretation is especially natural when optically thick cloud or haze layers dominate the radiative architecture, with their upper portions controlling much of the shortwave reflectivity and their lower portions contributing to broadband LTE thermal emission. Because $\alpha$ is defined at the top of the atmosphere, it should be interpreted as an emergent planetary quantity rather than as a direct measure of surface reflectivity alone. For Earth, Donohoe and Battisti \citep{donohoe2011atmospheric} estimate that approximately 88\% of the global-mean planetary albedo arises from atmospheric processes and about 12\% from surface-associated processes. The latter component is itself obtained through radiative decomposition and is therefore not independent of atmospheric transmission and model assumptions \citep{donohoe2011atmospheric}. From this perspective, Equation~\eqref{eq:1} provides a deliberately minimal closure in which stellar forcing enters through $I$ and the net shortwave-longwave coupling through the observable TOA parameter $\alpha$, without tunable coefficients.

A notable feature is the sensitivity near the theoretical bound $\alpha^\star = 1/r \approx 0.772$. Because Equation~\eqref{eq:1} contains the factor $(1-r\alpha)^{-1/4}$, small changes in $\alpha$ close to $\alpha^\star$ produce large shifts in $T_p$. Venus lies near this bound: taking the mean of two reported albedos, $0.76$ \citep{haus2016radiative} and $0.77$ from NASA fact sheets \citep{NASAfacts}, gives $\alpha = 0.765$ and a predicted $T_p = 742.3$\,K, in close agreement with the observed $737$\,K \citep{seiff1985models,NASAfacts}. Yet the two reported albedos alone yield a wide predicted range of $650$–$1024$\,K, underscoring that Equation~\eqref{eq:1} is highly sensitive near $\alpha^\star$. This sensitivity is a feature of the model which allows tighter albedo measurements to translate directly into sharper temperature constraints. Venus also illustrates the mechanism’s nuance. Its planet-encircling cloud deck is optically thick in the thermal infrared, such that the outgoing radiation to space is emitted primarily from cloud-top levels, while direct surface emission is confined to narrow spectral windows and constitutes only a small fraction of the total thermal flux \citep{crisp1986radiative}. Even this small escaping fraction is sufficient to establish radiative equilibrium and sustain the high surface temperature. The same cloud–haze “thermal mirror” that limits infrared escape also sets the proportionality between inner and Bond albedo, allowing the model to capture Venus within its domain of validity.

It is important to note that large surface temperatures are not inherently unphysical in strongly greenhouse-dominated regimes. Classical one-dimensional runaway greenhouse calculations for a steam atmosphere containing an ocean's worth of water yield surface temperatures exceeding 1500\,K \citep{kasting1988runaway}. While present-day Venus is not in such a steam-atmosphere runaway state, this example illustrates that very high temperatures can arise within physically consistent radiative--convective frameworks when infrared opacity becomes extreme. In the present model, Venus lies close to the limiting regime of the proposed closure, and its enhanced sensitivity reflects this structural feature rather than an inconsistency.

By contrast, away from the bound the response is weak. For Earth, with $\alpha=0.294\pm0.005$ \citep{stephens2015albedo}, the prediction $T_p=287.6\pm0.2$\,K agrees with the observed $288.2\pm0.5$\,K \citep{Hartmann2016Climate}. A $10\%$ perturbation of albedo would change $T_p$ only within $286.7$–$289.9$\,K. For Titan, with $\alpha=0.265\pm0.003$ \citep{creecy2021titan}, the prediction $92.5\pm0.04$\,K matches the observed surface temperature range $90.6$–$94$\,K \citep{mckay1991greenhouse,jennings2016surface}. A $10\%$ albedo change shifts $T_p$ only to $92.0$–$92.7$\,K. Thus Earth and Titan show stable predictions, while Venus illustrates extreme sensitivity near the bound.

For the gas giants, Equation~\eqref{eq:1} does not yield a solid-surface temperature but instead the temperature at the atmospheric level where absorbed solar radiation is balanced by thermal re-emission. Observationally, this corresponds to the onset of major species condensation and the transition to convective transport. On Jupiter, the predicted $133$\,K agrees with the measured $132.8$\,K at $0.5$\,bar \citep[Table~7]{seiff1998thermal}, consistent with NH$_3$ saturation and cloud formation \citep{atreya2005jupiter,kalogerakis2008coating,sromovsky2018composition,guillot2020storms1,guillot2020storms2,becker2020small}. On Saturn, the prediction $95.2\pm0.5$\,K is close to the Voyager occultation temperatures at specific pressures: $93.5$\,K at $0.275$\,bar and $94.8$\,K at $0.302$\,bar \citep{lindal1985atmosphere}, levels coinciding with the change in lapse rate and emergence of a lower convective haze layer \citep{hanel1983albedo}. This consistency reflects the fact that condensation levels mark the radiative–convective boundary, as noted in classical analyses of giant planet atmospheres \citep{atreya2005jupiter,guillot1995condensation}.

On Uranus, the prediction $66.2\pm0.1$\,K falls within the observed condensation range of $64.4$–$66.6$\,K, corresponding to ethane condensation near $14$\,mbar and evaporation near $600$\,mbar \citep{irwin2025uranus,lindal1987atmosphere,lunine1993atmospheres}. On Neptune, the prediction $52.4$\,K matches the observed tropopause (cold-trap) temperature $51$–$53$\,K where high-altitude hazes form and control the top-of-atmosphere albedo \citep{irwin2022hazy,lindal1990atmosphere,fletcher2010neptune}. In each case Equation~\eqref{eq:1} selects an outer atmospheric level associated with solar-forced radiative balance and major condensation. Internal heat sources remain important for the total emitted flux and for the deeper convective structure \citep{guillot1995condensation,christensen2020mechanisms}. The relevant point is that the characteristic temperature at this transition level may be identified from the radiative side through Equation~\eqref{eq:1}, while internal heating and the associated condensation physics determine the same region from below. These are therefore best viewed as complementary descriptions of the same atmospheric transition and should converge on the same characteristic temperature. Titan provides a bridge between terrestrial and giant planets. Its surface temperature 
$90.6$--$94$\,K \citep{mckay1991greenhouse,jennings2016surface} is reproduced by the closure 
relation, and similar temperatures occur again in the lower stratosphere near 
$\sim$30\,mbar \citep{fulchignoni2005titan}, where in-situ measurements and microphysical 
models indicate the presence of a condensate haze layer formed by the downward transport and 
growth of photochemical aerosols \citep{tomasko2008model,lavvas2008coupling}, echoing the 
condensation-level interpretation seen in the giants.

\begin{figure}[t]
\centering
\includegraphics[width=0.5\textwidth]{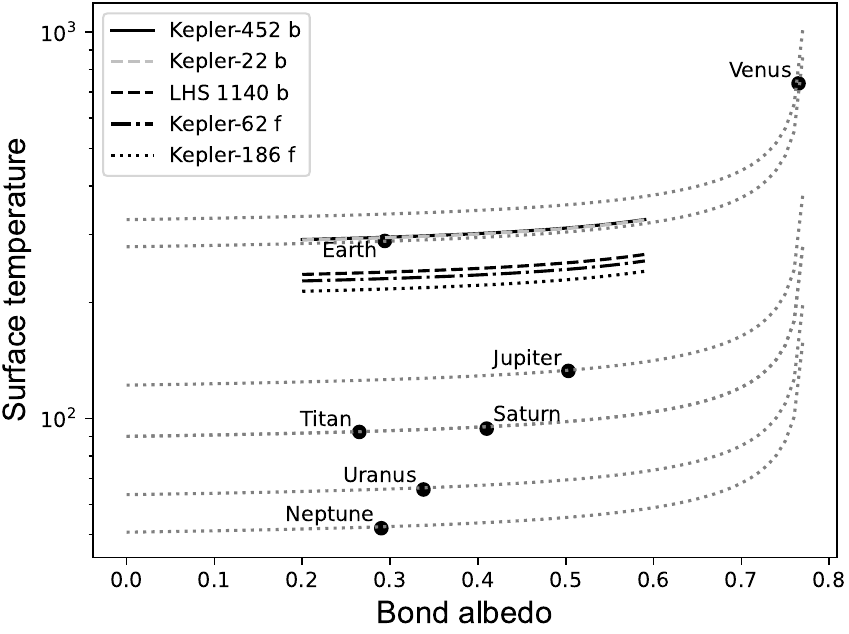}
\caption{Predicted equilibrium surface or reference-level temperatures from Equation~\eqref{eq:1} as a function of planetary Bond albedo. Data are shown for Solar System planets and moons with substantial atmospheres, together with selected habitable-zone exoplanets. For exoplanets, shaded ranges indicate predicted surface temperatures for $\alpha \in [0.2,0.6]$.}
\label{fig:4}
\end{figure}

Figure~\ref{fig:4} summarizes the predictions of Equation~\eqref{eq:1} for Solar System bodies with substantial atmospheres, alongside several well-characterized exoplanets in the habitable zone. The vertical axis shows the observed or reference temperature, while the horizontal axis shows the measured or estimated Bond albedo. The curve represents the prediction of Equation~\eqref{eq:1}, and the planetary data points fall remarkably close to it across nearly three orders of magnitude in solar irradiance. For Venus, Earth, Titan, Jupiter, Saturn, Uranus, and Neptune, the agreement is within uncertainties of the measured albedo $\alpha$ and surface temperature $T_p$.  

Another possible application of our results is in exoplanet studies, where large-scale analyses of planetary populations and habitability are routinely performed \citep{kopparapu2013habitable,jiang2024analysis}. Using the published stellar and orbital parameters to set the insolation $S$ \citep{jenkins2015kepler452b,borucki2012kepler22b,dittmann2017lhs1140b,ment2019lhs1140system,borucki2013kepler62f,quintana2014kepler186f}, we evaluate the surface temperature for exoplanets over a plausible Bond-albedo range $\alpha \in [0.2,0.6]$. Over this interval, our model predicts temperate conditions for Kepler-452\,b and Kepler-22\,b, with $T\simeq 291$--$330$\,K and $T\simeq 290$--$329$\,K, respectively. By contrast, LHS~1140\,b, Kepler-62\,f, and Kepler-186\,f remain colder across the same sweep, with predicted ranges below the freezing point of water: $236$--$268$\,K, $227$--$258$\,K, and $213$--$242$\,K, respectively. In the literature, these cooler planets are sometimes discussed as potentially habitable under strong greenhouse scenarios \citep{kopparapu2013habitable,shields2016habitability}. Similarly, our purely radiative--geometric framework places their mean equilibrium below $273$\,K, highlighting the need for additional mechanisms to raise surface temperatures above freezing.

\subsection{Scope and Limitations}

Earth provides the most stringent test: with a precisely measured albedo and surface temperature, Equation~\eqref{eq:1} reproduces the observed mean within $0.2$\,K. This level of agreement indicates that, for well-characterized systems, the relation captures a dominant component of planetary energy balance using only directly measurable quantities.

At the same time, the present framework is intentionally formulated as a coarse-grained closure rather than a detailed physical model. Its role is analogous to that of classical effective descriptions, but with a different organizing principle: instead of introducing independent parameters for longwave radiative effects, it tests whether the observable top-of-atmosphere Bond albedo alone can constrain both shortwave reflection and longwave return through a single proportionality. The relation $\beta = r\alpha$ should therefore be understood as a closure hypothesis supported by empirical consistency across planetary environments and given a geometric rationale in Section 2 and Appendix~A, rather than as a derivation from first-principles radiative transfer.

In this sense, the present approach complements, rather than replaces, standard atmospheric modeling. Radiative--convective and general circulation models remain essential for resolving vertical structure, dynamics, and feedbacks. The contribution of Equation~\eqref{eq:1} is instead to identify a reduced constraint at the level of global energy balance, against which more detailed models may be compared. Its utility lies precisely in the fact that it operates with minimal inputs $(I,\alpha)$ and no adjustable parameters.

The contrast with classical effective models is instructive. In one-layer and multi-layer formulations, the atmospheric effect is represented through parameters such as effective emissivity and the number of layers, which must be specified or fitted for each planet. This becomes particularly evident in extreme regimes. Because the emissivity is bounded by unity, the one-layer model cannot reproduce Venus without introducing additional layers, thereby increasing the number of adjustable parameters. By comparison, Equation~\eqref{eq:1} captures Venus as a limiting case near $\alpha^\star = 1/r$, where the system becomes highly sensitive to albedo. This sensitivity is reflected in the strong dependence of the predicted temperature on small variations in $\alpha$, and is consistent with the observed behavior of optically thick cloud decks that strongly regulate infrared escape \cite{kasting1988runaway}.

An alternative class of simplified approaches infers surface temperature from the environmental lapse rate, by identifying an effective emission level and extrapolating downward along an assumed temperature gradient \citep{hansen1981climate,wallace2006atmospheric}. These methods provide a useful low-dimensional description of atmospheric structure and can yield reasonable estimates when the lapse rate is well constrained. However, they operate by prescribing the vertical temperature profile rather than by explicitly encoding the global radiative energy balance. As a result, the inferred surface temperature depends on assumptions about the location of the emission level and the form of the lapse rate, which may vary across planetary environments. In this sense, lapse-rate-based estimates are complementary to the present framework: they describe how temperature varies within the atmosphere, whereas Equation~\eqref{eq:1} constrains the overall radiative equilibrium using top-of-atmosphere observables. Together, these perspectives highlight different aspects of the same system, but they rely on distinct inputs and assumptions.

The applicability of the framework is also clarified by boundary cases. Mars provides a natural test of the thin-atmosphere limit: despite having a CO$_2$ atmosphere and episodic cloud formation, its low surface pressure of approximately 6\,mbar \citep{NASAfacts}, less than 1\% of Earth’s, and weak radiative coupling lead to surface temperatures close to the blackbody estimate. Its atmosphere is therefore optically thin in the infrared compared to Earth or Venus, and in this regime the effective return fraction $\beta$ is small, so Equation~\eqref{eq:1} reduces toward the $\beta = 0$ limit, namely the standard radiative equilibrium expression. This behavior is consistent with the interpretation that the relation applies primarily to atmospheres with optically significant cloud or haze layers capable of sustaining substantial longwave radiative return.

For the giant planets, the interpretation requires additional care. Equation~\eqref{eq:1} does not predict a solid surface temperature but instead identifies an outer atmospheric level associated with radiative balance and major condensation. Internal heat sources remain important for the total emitted flux and for the deeper convective structure, but the characteristic temperature at the transition region can be identified from the radiative side through Equation~\eqref{eq:1}, while internal heating and associated condensation physics determine the same region from below. These should therefore be regarded as complementary constraints acting on a shared atmospheric transition.

Uncertainty propagation has been incorporated explicitly through the dependence on $\alpha$, which is the dominant observational input. As illustrated in Table~\ref{table:1} and Figure~\ref{fig:min_r}, uncertainties in albedo translate directly into temperature uncertainties, with sensitivity increasing near the theoretical bound $\alpha^\star$. This behavior is not incidental but reflects the structure of Equation~\eqref{eq:1}, and it highlights the importance of accurate albedo measurements for testing the model. To facilitate reproducibility, the analysis underlying the figures has been implemented in code and can be directly evaluated for different planetary datasets.

Finally, the geometric construction introduced in Section 2 is not intended to impose a global symmetry on planetary atmospheres, but to provide a local, beam-aligned representation that motivates the emergence of a near-constant proportionality factor. Appendix~A develops this point in detail, showing that the two-dimensional construction is the thin-slab reduction of a three-dimensional description and that it is consistent, at a general level, with standard radiative laws.

Taken together, these considerations define the scope of the model: it is a minimal, testable closure at the level of global energy balance, applicable to planets with substantial atmospheres and significant radiative coupling, and intended as a complementary constraint alongside more detailed physical models.

\section{Conclusions}

This study shows that planetary surface temperatures are not unconstrained consequences of atmospheric complexity alone, but can be organized by a compact closure linking irradiance, albedo, and temperature through a geometric coefficient. In this sense, the framework points to a previously under-emphasized structural relation within planetary energy balance. Its apparent success across both rocky planets and gas giants suggests that, at planetary scale, an important part of climate regulation may be captured by a geometric constraint operating on bulk radiative quantities, rather than only by the accumulation of model-specific parameters.

Within this perspective, the Bond albedo $\alpha$ assumes a more central role than it is often given in simplified discussions of planetary temperature. As a top-of-atmosphere quantity, $\alpha$ is not merely a descriptive measure of reflected stellar radiation, but an integrated radiative diagnostic of the processes that shape the partitioning between absorbed shortwave energy and its return to space. The present results suggest that these processes, and their connection to longwave radiative return, deserve more explicit attention in the theoretical interpretation of planetary climate. In the proposed closure, this connection is represented through the relation between $\alpha$ and $\beta$, so that the large-scale structures governing shortwave reflection are treated as effectively coupled to those governing downward thermal re-emission. The proportionality is expressed by a geometric factor associated with the parabolic constant, yielding a reduced description in which the radiative influence of atmospheric structure enters through a single effective relation between $\alpha$ and $\beta$.

The framework correspondingly spans diverse planetary regimes: it reproduces the climates of Earth, Venus, and Titan, and captures the condensation-level temperatures of the gas giants. Its scope also extends to exoplanets, where it offers first-order estimates of equilibrium surface conditions and helps identify cases in which additional heating may be required to sustain liquid water. Relative to radiative-convective or multi-layer approaches, the point is not to replace detailed modeling where such modeling is needed, but to show that a substantial fraction of planetary temperature structure may already be constrained at the level of TOA radiative observables. From this perspective, Bond albedo should be regarded not as a secondary input parameter, but as a fundamental quantity whose physical interpretation, and whose relation to atmospheric thermal return, merits closer attention in the broader climate literature.

The model limitations are that it does not apply to bodies without substantial atmospheres (e.g., Mercury, the Moon, Mars, Triton), where large diurnal or seasonal swings preclude a well-defined mean temperature \citep{vasavada1999near,leovy2001weather,read2015physics,trafton1984large}, nor does it capture internal heat sources in giant planets below the radiative balance level. It is not a substitute for line-by-line radiative transfer or full general circulation models. Rather, it indicates that the re-radiation parameter in textbook energy-balance models may be constrained by $\alpha$ in cloud-rich regimes, because both quantities can be shaped by the same large-scale atmospheric structures and are therefore not necessarily independent degrees of freedom \citep{wallace2006atmospheric}. Within scope, however, the framework is universal, testable, and falsifiable: it predicts observed temperatures across diverse planetary environments using only $(I,\alpha)$ and the geometric constant $r$. Mars illustrates the boundary case: despite a thin CO$_2$ atmosphere and occasional clouds, its low pressure and transient cloud cover render the thermal mirror mechanism ineffective, leaving its mean surface temperature close to the blackbody estimate. This is consistent with the restriction to planets possessing optically thick, globally significant cloud or haze layers.

The broader implication is that planetary climate regulation can, to a significant extent, be organized through a geometric relation among bulk radiative quantities. From this viewpoint, the Bond albedo should be interpreted not merely as an observational descriptor, but as a physically consequential quantity that encodes the radiative processes linking shortwave reflection to atmospheric thermal return. This perspective provides a unifying lens for comparative planetology, clarifies the role of atmospheric structure in planetary energy balance, and offers a useful framework for prioritizing exoplanet habitability assessments. As future observations place tighter constraints on exoplanet albedos and on the atmospheric properties that shape them, the present framework may enable correspondingly improved estimates of surface temperature and habitability.

\section*{Funding}
This research received no external funding.

\section*{Institutional Review Board Statement}
Not applicable.

\section*{Informed Consent Statement}
Not applicable.

\section*{Data Availability Statement}
No new data were created or analyzed in this study. All data used are drawn from published sources cited in the manuscript. The code used in the analysis and to produce the figures is available here: \url{https://zenodo.org/records/19640331}

\section*{Conflicts of Interest}
The author declares that he has no known competing financial interests or personal relationships that could have appeared to influence the work reported in this paper.

\section*{Use of AI and AI-assisted Technologies}
Generative AI and AI-assisted tools were used solely for language editing, phrasing, and editorial assistance during the preparation of this manuscript. The author reviewed and edited all AI-assisted output and takes full responsibility for the content of the publication.

\appendix

\section*{Appendix A. Geometric Justification of the Inner-Albedo Construction}

This appendix develops the geometric construction used in Section 2 and places it in a corresponding three-dimensional setting. The central point is that the parabolic geometry arises naturally in a beam-aligned local cross-section, and that the same construction extends consistently to a three-dimensional description through a decomposition into thin overlapping slabs. In this way, the two-dimensional argument in the main text is interpreted as a local geometric building block rather than as a standalone global model.

\subsection*{Appendix A.1. Local Geometric Relation used in Figure~4}

This subsection develops the local approximation used in Section 2 to relate the short mirror segment to the corresponding parabolic segment. In the notation of Figure~4, the focus $F$ has coordinate $x_F=0$, while the coordinate of $G$ is
\begin{equation}
x_G
=
\frac{R}{R+h}x_B
+
\frac{y_B}{R+h}\sqrt{x_B^2+y_B^2-R^2},
\end{equation}
where $R$ is the radius of $S$ and $h$ is the distance between $S$ and $M$. Under the assumptions
\begin{equation}
h \ll AB \ll R,
\end{equation}
one obtains
\begin{equation}
x_G \simeq 2x_B.
\end{equation}
Since $x_A=-x_B$, this gives
\begin{equation}
AB=2x_B,
\end{equation}
and therefore
\begin{equation}
FG \simeq AB.
\end{equation}
Because the circular segment $AB$ is short compared with the planetary radius, its arc length satisfies
\begin{equation}
l_1 \simeq AB \simeq FG.
\end{equation}
This is the local geometric identification used in Section 2.

The corresponding parabolic segment has length $L_1$, and the relevant ratio is the parabolic constant
\begin{equation}
p=\sqrt{2}+\log(1+\sqrt{2}),
\end{equation}
so that
\begin{equation}
\frac{L_1}{l_1}\simeq p.
\end{equation}
Substituting this into the two-dimensional conservation law
\begin{equation}
l_1(I_0+I_1)=L_1 I_2
\end{equation}
gives
\begin{equation}
I_1=
\left(\frac{L_1}{l_1}-1\right)I_0
\simeq
(p-1)I_0.
\end{equation}
Summing over the reflecting segments then yields the effective inner-albedo relation
\begin{equation}
\beta=r\alpha,
\qquad
r=p-1.
\end{equation}
Thus the coefficient used in the main text is a local geometric factor derived from the ratio between a finite parabolic arc and its corresponding chord.

\subsection*{Appendix A.2. Beam-aligned Geometry, Focal-angle Parametrization, and the Three-dimensional Extension}

This subsection contrasts the present construction with ordinary isotropic point-source geometry and develops the corresponding beam-aligned three-dimensional interpretation. If radiation emitted from a point were distributed purely by distance from that point, then the natural level sets would be
\begin{equation}
r=\mathrm{const.},
\end{equation}
namely circles in two dimensions and spheres in three dimensions. This is the familiar radial picture. In the present setting, however, the outgoing beam singles out a preferred propagation direction, so the relevant geometry is not purely radial. Instead, one should measure distance relative to the beam axis.

Accordingly, place the effective emitting point at the origin and let the positive $z$-axis point upward from the surface into the atmosphere. In the beam-aligned two-dimensional cross-section, let
\begin{equation}
r=\sqrt{x^2+z^2},
\qquad
z=r\cos\theta,
\end{equation}
where $\theta$ is the angle measured from the positive $z$-axis. The natural beam-aligned family is then
\begin{equation}
r+z=c,
\label{eq:rzc}
\end{equation}
where $c>0$ is a constant. Geometrically, this is precisely the focus--directrix form of a parabola: the distance from a point $(x,z)$ to the focus at the origin is $r$, while its distance to the horizontal directrix $z=c$ is $c-z$, so the defining relation
\begin{equation}
r=c-z
\end{equation}
is equivalent to the beam-aligned relation above.

The parameter $c$ therefore labels a family of confocal fronts; it is the height of the corresponding directrix above the focus, and it determines the position of the vertex. In Cartesian coordinates, the relation becomes
\begin{equation}
\sqrt{x^2+z^2}+z=c
\quad \Longleftrightarrow \quad
x^2=-2cz+c^2,
\end{equation}
which is a parabola with focus at the origin and vertex at
\begin{equation}
\left(0,\frac{c}{2}\right).
\end{equation}
Thus the relevant parabolic front has its tip above the effective focus and is concave downward, consistent with the geometry used in Section 2.

This also clarifies the role of angle. Solving the beam-aligned relation for $r$ gives
\begin{equation}
r(\theta;c)=\frac{c}{1+\cos\theta}.
\label{eq:polarparabola}
\end{equation}
Thus each front may equally well be parametrized by the focal angle $\theta$. Since $\theta$ is measured from the positive $z$-axis, the physically relevant upward segment is naturally centered at $\theta=0$. In practice, one considers only the finite segment intercepted by the relevant beam, corresponding to a finite angular window
\begin{equation}
-\theta_0 \le \theta \le \theta_0 .
\end{equation}
The quantity $\theta_0$ is therefore the half-angle of the beam-selected sector as seen from the effective focus. In this way, the finite parabolic segment used in the main text is the image of a finite angular sector of the beam-aligned geometry.

The corresponding three-dimensional construction is obtained in exactly the same way. Let $\theta$ be the polar angle from the positive $z$-axis and let
\begin{equation}
r=\sqrt{x^2+y^2+z^2},
\qquad
z=r\cos\theta.
\end{equation}
Then the same beam-aligned family is
\begin{equation}
r+z=c,
\end{equation}
and its level sets are
\[
\begin{aligned}
r+z=c
\quad \Longleftrightarrow \quad
\sqrt{x^2+y^2+z^2}+z&=c\\
x^2+y^2&=-2cz+c^2.
\end{aligned}
\]
These are paraboloids of revolution with common focus at the origin and vertex above the focus. The two-dimensional parabola used in Section 2 is therefore the beam-aligned cross-section of the corresponding three-dimensional paraboloidal geometry.

The distinction between the two-dimensional and three-dimensional descriptions is then straightforward. In full three dimensions one transports power across a paraboloidal patch cut out by a fixed solid-angle element $d\Omega$, so the appropriate conservation law is area-based:
\begin{equation}
I_{3D}(c)\,A(c;d\Omega)=\mathrm{const.}
\end{equation}
For fixed solid angle, the linear size of the patch grows like $c$, hence its area grows like
\begin{equation}
A(c;d\Omega)\propto c^2,
\end{equation}
and therefore
\begin{equation}
I_{3D}(c)\propto c^{-2}.
\end{equation}
Thus the usual inverse-square behavior is recovered in the beam-aligned three-dimensional picture.

The two-dimensional law used in Section 2 arises by restricting attention not to a full solid-angle patch, but to a thin azimuthal slab of width $\Delta\phi$. Such a slab is the three-dimensional realization of a beam-aligned cross-section. At parameter value $c$, its transverse physical width scales as
\begin{equation}
w_\perp(c)\sim c\,\Delta\phi.
\end{equation}
If $L(c;\theta_0)$ denotes the length of the parabolic front segment corresponding to the angular window $[-\theta_0,\theta_0]$, then the power carried by the slab is approximately
\begin{equation}
P_{\mathrm{slab}}
\approx
I_{3D}(c)\,L(c;\theta_0)\,w_\perp(c)
\approx
I_{3D}(c)\,L(c;\theta_0)\,c\,\Delta\phi.
\end{equation}
This motivates the reduced slab intensity
\begin{equation}
I_{\mathrm{slab}}(c):=I_{3D}(c)\,c\,\Delta\phi,
\end{equation}
for which the conservation law becomes purely two-dimensional:
\begin{equation}
I_{\mathrm{slab}}(c)\,L(c;\theta_0)=\mathrm{const.}
\end{equation}
Since, for fixed angular window $[-\theta_0,\theta_0]$, the front length scales linearly with $c$,
\begin{equation}
L(c;\theta_0)\propto c,
\end{equation}
it follows that
\begin{equation}
I_{\mathrm{slab}}(c)\propto c^{-1}.
\end{equation}
Dividing once more by the transverse width factor recovers
\begin{equation}
I_{3D}(c)\propto c^{-2}.
\end{equation}

In this way, the two-dimensional construction of Section 2 is seen to be the thin-slab reduction of a fully three-dimensional beam-aligned geometry. The short parabolic cylinders invoked in the main text are precisely the local three-dimensional fronts obtained by extending, over a small azimuthal width, the beam-aligned parabolic cross-section defined by $r+z=c$.

\subsection*{Appendix A.3. Finite Emitting Regions and Overlap Statistics}

This subsection formalizes the decomposition of a finite emitting region into thin beam-aligned slabs and relates the uncovered fraction to an effective geometric optical depth.

For a finite emitting region, the field may be represented as a superposition of many thin beam-aligned slabs. Each slab carries a local parabolic-cylindrical front, and the full radiative field is obtained by summing over their overlapping contributions.

To formalize this, consider the disk
\begin{equation}
D_R=\{(x,y)\in\mathbb{R}^2:x^2+y^2\le R^2\},
\end{equation}
where \(R>0\) is the disk radius. A strip is specified by an orientation \(\theta\in[0,\pi)\), an offset \(s\in[-R,R]\), and a width \(w>0\). Writing
\begin{equation}
n_\theta=(\cos\theta,\sin\theta),
\end{equation}
the corresponding strip is
\begin{equation}
S(\theta,s)=\{x\in\mathbb{R}^2:\ |x\cdot n_\theta-s|\le w/2\},
\end{equation}
and its portion inside the disk is
\begin{equation}
T(\theta,s)=D_R\cap S(\theta,s).
\end{equation}
We consider \(N\) such strips,
\begin{equation}
T_i:=T(\theta_i,s_i), \qquad i=1,\dots,N,
\end{equation}
with \(\theta_i\) independent and uniformly distributed on \([0,\pi)\), and \(s_i\) independent and uniformly distributed on \([-R,R]\).

Fix a point \(x\in D_R\). For any given strip, the condition \(x\in T_i\) is
\begin{equation}
|x\cdot n_{\theta_i}-s_i|\le w/2.
\end{equation}
Since \(s_i\) is uniform on \([-R,R]\), the probability that a given strip covers \(x\) is
\begin{equation}
p=\frac{w}{2R}.
\end{equation}
This quantity is dimensionless and independent of both \(x\) and \(\theta_i\). Hence, for any \(m\) distinct strips,
\begin{equation}
\mathbb{P}(x\in T_{i_1}\cap\cdots\cap T_{i_m})=p^m.
\end{equation}
Integrating over the disk gives
\begin{equation}
\mathbb{E}[|T_{i_1}\cap\cdots\cap T_{i_m}|]=\pi R^2 p^m.
\end{equation}
Therefore the expected covered area of the union
\begin{equation}
U_N:=\bigcup_{i=1}^N T_i
\end{equation}
is
\begin{align}
\mathbb{E}[|U_N|]
&=
\pi R^2\sum_{m=1}^{N}(-1)^{m+1}\binom{N}{m}\left(\frac{w}{2R}\right)^m \\
&=
\pi R^2\left[1-\left(1-\frac{w}{2R}\right)^N\right].
\end{align}
Hence the expected covered-area fraction is
\begin{equation}
\frac{\mathbb{E}[|U_N|]}{\pi R^2}
=
1-\left(1-\frac{w}{2R}\right)^N.
\end{equation}

For equal-width strips, it is convenient to introduce the dimensionless overlap parameter
\begin{equation}
\mu:=\frac{Nw}{2R}.
\end{equation}
Taking the continuum limit
\begin{equation}
N\to\infty,\qquad w\to 0,\qquad \mu=\frac{Nw}{2R}\ \text{fixed},
\end{equation}
one obtains
\begin{equation}
\left(1-\frac{w}{2R}\right)^N\to e^{-\mu}.
\end{equation}
Therefore
\begin{equation}
\frac{\mathbb{E}[|U_N|]}{\pi R^2}\to 1-e^{-\mu},
\end{equation}
and the complementary fraction tends to
\begin{equation}
e^{-\mu}.
\end{equation}
Thus \(\mu\) is the natural dimensionless measure of overlap in the strip decomposition. For variable strip widths \(w_i\), the corresponding form is
\begin{equation}
\mu=\frac{1}{2R}\sum_i w_i.
\end{equation}
In this sense, the quantity \(\sum_i w_i\) is the effective covered transverse measure.

As a distinct application of the slab decomposition, one may recover the Beer--Lambert law. To do so, introduce a fixed inverse length scale \(\gamma\), analogous in dimension to an attenuation coefficient such as \(\kappa\), and write
\begin{equation}
Nw=\gamma A,
\end{equation}
where \(A\) is an effective covered area. Then
\begin{equation}
\mu=\frac{\gamma A}{2R},
\end{equation}
which is manifestly dimensionless.

Now let the front propagate through a uniform medium over a path length \(L\), and let \(R(L)\) and \(A(L)\) denote, respectively, the transverse radius of the front and the corresponding effective covered area. For ordinary expanding wavefronts one expects
\begin{equation}
R(L)\propto L,
\end{equation}
and therefore
\begin{equation}
A(L)\propto L^2.
\end{equation}
It follows that
\begin{equation}
\mu(L)=\frac{\gamma A(L)}{2R(L)}\propto \gamma L,
\end{equation}
so the overlap parameter grows linearly with path length while remaining dimensionless, as required for an optical-depth-like quantity.

If \(I_0\) is the initial intensity and \(I(L)\) the intensity after propagation over a distance \(L\), then the exponential factor takes the form
\begin{equation}
\frac{I(L)}{I_0}=e^{-\tau(L)},
\end{equation}
with
\begin{equation}
\tau(L)\leftrightarrow \mu(L).
\end{equation}
In this interpretation, \(\mu(L)\) plays the role of an effective geometric optical depth, while \(\gamma\) plays the role of the corresponding attenuation coefficient. This derivation is not used in the main text; its purpose is only to show that the same slab-wise decomposition also reproduces the exponential structure of a standard radiation law, thereby illustrating the generality of the construction.

\subsection*{Appendix A.4. Summary}

The appendix establishes four points. First, the coefficient used in Section 2 is obtained from the ratio between a finite parabolic arc and its corresponding chord, giving $r=p-1$. Second, the parabola appearing in the main text is the natural beam-aligned cross-section of a three-dimensional paraboloidal geometry. Third, the two-dimensional conservation law is the thin-slab reduction of the corresponding three-dimensional inverse-square radiative scaling. Fourth, when a finite emitting region is decomposed into many thin overlapping slabs, the uncovered fraction converges to $e^{-\mu}$, which is mathematically identical to the Beer--Lambert transmission law $e^{-\tau}$, with $\mu$ acting as an effective geometric optical depth.

Taken together, these points show how the local parabolic construction used in Section 2 fits into a broader three-dimensional and attenuation-consistent framework.

\bibliographystyle{apsrev4-2}
\bibliography{apssamp}

\end{document}